\documentclass{article} % For LaTeX2e
\usepackage{iclr2023_conference,times}

\usepackage{hyperref}
\usepackage{url}
\usepackage{graphicx}

\title{Long-lead forecasts of wintertime air stagnation index in southern China using oceanic memory effects} %with LSTMs  Statistical predictability
\author{Chenhong Zhou \\
Department of Computer Science \\ Hong Kong Baptist University \\
\texttt{20482795@life.hkbu.edu.hk} \\
\And
Xiaorui Zhang \&  Meng Gao \\
Department of Geography \\ Hong Kong Baptist University \\
\And
Shanshan Liu \\
University of Science and \\ Technology of China 
\And
Yike Guo \\
The Hong Kong University \\ of Science and Technology
\And
Jie Chen \thanks{Corresponding author.} \\
Department of Computer Science \\ Hong Kong Baptist University \\
\texttt{chenjie@comp.hkbu.edu.hk}
}

\iclrfinalcopy  % Uncomment for camera-ready version, but NOT for submission.

\begin{document}

\maketitle

\begin{abstract} 
Stagnant weather condition is one of the major contributors to air pollution as it is favorable for the formation and accumulation of pollutants. To measure the atmosphere’s ability to dilute air pollutants, Air Stagnation Index (ASI) has been introduced as an important meteorological index. Therefore, making long-lead ASI forecasts is vital to make plans in advance for air quality management. In this study, we found that autumn Niño indices derived from sea surface temperature (SST) anomalies show a negative correlation with wintertime ASI in southern China, offering prospects for a prewinter forecast. We developed an LSTM-based model to predict the future wintertime ASI. Results demonstrated that multivariate inputs (past ASI and Niño indices) achieve better forecast performance than univariate input (only past ASI). The model achieves a correlation coefficient of 0.778 between the actual and predicted ASI, exhibiting a high degree of consistency.

\end{abstract}

% \vspace{0.2cm}

\section{Introduction}
Air pollution has become a focus problem which various countries are concerned about. The growth of atmospheric contaminants damages vegetation and crops and even is highly related to some serious human health diseases \citep{kampa2008human, al2005improving, bai2018air, Masood2021review}. Apart from direct emissions of air pollutants, meteorological conditions are also important to the accumulation and dispersion of pollutants. Air stagnation occurs usually with descending air, low wind speeds, low precipitation, and the compressed boundary layer, which is favorable for the formation and accumulation of atmospheric contaminants \citep{wu2017atmospheric, gao2019seasonal}. Hence, the Air Stagnation Index (ASI) has been proposed to assess the atmosphere's ability to dilute air pollutants \citep{horton2012response, horton2014occurrence, Huang2018climatological}. Studies have shown that air stagnation highly correlates with the concentrations of air pollutants \citep{liao2006role, Huang2018climatological}. Therefore, accurate forecasts of ASI are important and valuable for managing air quality and enabling advance planning.

However, most existing works usually forecast the next several hours or days of air pollution levels in advance by using common meteorological variables (e.g., wind speed, wind direction, humidity, temperature, and rainfall), ground-level observations, and satellite data, etc  \citep{al2005improving, bai2018air, li2011study, harishkumar2020forecasting, ham2019deep,chae2021pm10,zhang2020deep, xiao2020improved, chen2021lstm, kurt2010forecasting,  ham2021unified, Chang2020lstm, Wu20191daily}. At present, long-lead seasonal or multi-year air pollution forecasts are still under exploration, because the ability of long-lead forecasts is highly dependent on finding strongly correlated climate factors and appropriate forecast algorithms. Previous studies have linked the long-lead El Niño/Southern Oscillation (ENSO) \citep{ham2019deep, ham2021unified}, Indian wintertime aerosol pollution \citep{gao2019seasonal} and wintertime PM$_{2.5}$ concentrations over East Asia \citep{jeong2021statistical} with ocean memory effects. They are based on the fact that sea surface temperatures (SST) vary slowly and the presence of decadal oceanic variations as well as their coupling to the atmosphere would modulate the interannual variability of climate change and air pollution. Given these linkages, we aim to investigate the relationship between ocean memory effects and ASI forecasts.

\paragraph{Contributions} We first use the meteorological variables from the ERA5 reanalysis dataset \citep{Hersbach2020ERA5} to generate a long-term ASI dataset over China during a long period from 1950 to 2020. To investigate the feasibility of long-lead ASI forecasts, we analyze the correlations between ASI and ENSO-related indices (Niño 1+2, Niño 3, Niño 3.4, Niño 4) calculated from SST anomalies. The correlation map shows that Niño indices are negatively associated with seasonal variations of ASI in southern China. Furthermore, we develop the Long Short-Term Memory (LSTM) using Niño indices as predictors to obtain skillful long-lead forecasts of wintertime ASI.

\paragraph{Pathways to Climate Impact } We show the statistical predictability of wintertime ASI in southern China using machine learning with ocean memory effects as predictors.  Our encouraging results suggest that SST patterns play an important role in long-lead forecasts of wintertime ASI, which is useful for analyzing the impact of climate patterns on air pollution.

\begin{figure}[!t]
  \centering
  \includegraphics[width=\linewidth]{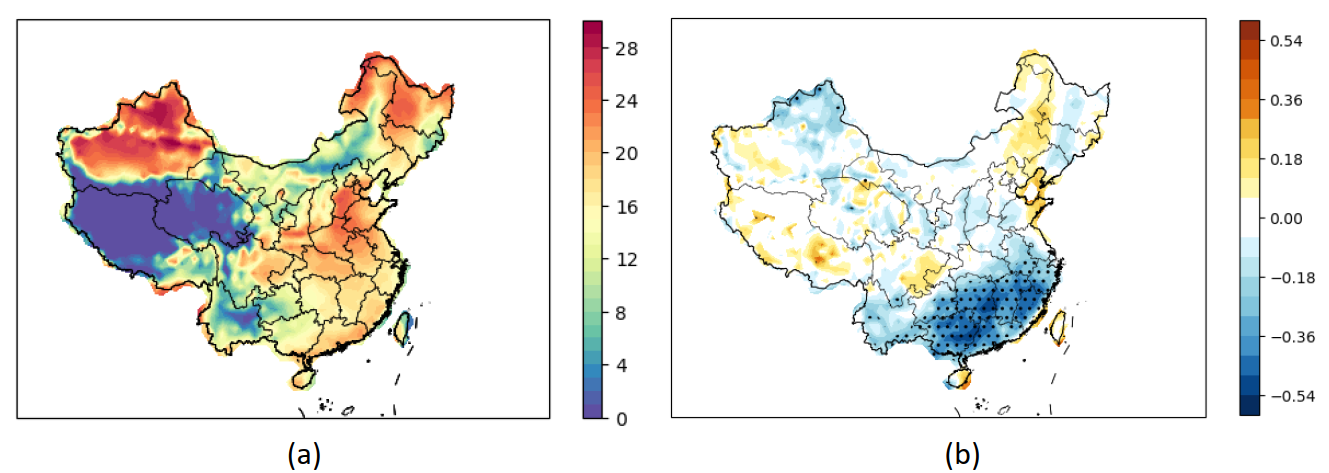}
  \caption{(a) Multi-year (1950–2020) wintertime monthly mean (From December of one year to February of the following year) ASI distribution across China. (b) Correlation coefﬁcients between autumn Niño 3.4 and wintertime ASI over China, where black dots denote  significance (p \textless 0.05).
  }
  \label{fig1}
\end{figure}

\vspace{0.5cm}

\section{Methodology}

\paragraph{Data}  There is no available ready-made ASI dataset across China, so we first generate a long-term ASI dataset during the period of 1950 to 2020 using the meteorological variables from ERA5 reanalysis dataset \citep{Hersbach2020ERA5}.  Following the computation process of 
ASI in \citep{garrido2021assessing}, we downloaded these meteorological variables from the ERA5 reanalysis dataset, including convective available potential energy, boundary layer height, convective inhibition, daily wind speed at different heights, and daily accumulated precipitation, etc, to compute the ASI across China. 
According to the definitions in \citep{garrido2021assessing}, we determine whether air stagnation occurs by judging whether the meteorological condition meets some predefined thresholds of daily meteorological fields. Hence, ASI on one day is a  binary value, i.e. stagnant/non-stagnant, and the monthly average ASI can be easily obtained by accumulating the values of ASI in one month. A long-term ASI dataset is finally generated and we display the multi-year wintertime monthly mean ASI distribution across China in Figure~\ref{fig1} (a).

Four ENSO-related indices: Niño 1+2, Niño 3, Niño 3.4, and Niño 4, derived from SST anomalies are downloaded from \url{https://www.cpc.ncep.noaa.gov/data/indices/}. 
To prove the feasibility of predicting wintertime ASI using Niño indices, we need to investigate the relationships between them before constructing a forecast model. More specifically, we calculate the Pearson correlation coefﬁcients between wintertime ASI across China and autumn Niño indices. Figure~\ref{fig1} (b) shows the correlation coefﬁcient map where black dots mean statistically signiﬁcant relationships. It is observed that autumn Niño 3.4 index shows strong negative correlations with wintertime ASI in southern China. The negative correlations between Niño 3.4 index and air pollution in southern China have also been discussed in \citep{Zhao2018effects, Cheng2019climate}. Therefore, we select the statistically signiﬁcant region, i.e., seven provinces in southern China, including Guizhou, Hunan, Jiangxi, Fujian, Zhejiang, Guangxi, and Guangdong. Then we calculate the average monthly ASI of this region and obtain the spatially average wintertime ASI time-series data. The Pearson correlation coefficients between this ASI time-series data and autumn Niño 1+2, Niño 3, Niño 3.4, and Niño 4 are -0.42, -0.50, -0.46, and -0.53, respectively. These correlations suggest that making seasonal or even interannual predictions of wintertime ASI is possible.

\paragraph{Model and Experimental Setup} 
We develop a forecast model by exploiting the LSTM cell which is good at capturing the long-term dependency of time-series data \citep{hochreiter1997long}. 
This LSTM-based model is composed of one input layer, two-layer LSTMs (each LSTM layer with 32 hidden states), and one output layer. To explore the effects of September-October-November (SON) Niño indices on foreseeing future December–January–February (DJF) ASI over southern China, we make comparisons between univariate input and multivariate inputs. The univariate input refers to past DJF ASI, while multivariate inputs refer to past DJF ASI plus SON Niño indices. The input sequence length is denoted as $k$, which means the input sequence from time $\tau-k+1$ months to time $\tau$ (in months). The output layer is a fully connected (FC) layer whose variable is the future ASI at time $t$ ($t >= \tau +1$). The training period is from 1950 to 1999 used to train the LSTM model and the period for validating the forecast skill is from 2000 to 2020.  Adam optimizer \citep{kingma2014adam} with a learning rate of 0.001 and 0.01 is used for univariate and multivariate inputs respectively, and mean-squared error (MSE) is chosen as our loss function.

\begin{figure}[!t]
  \centering
  \includegraphics[width=\linewidth]{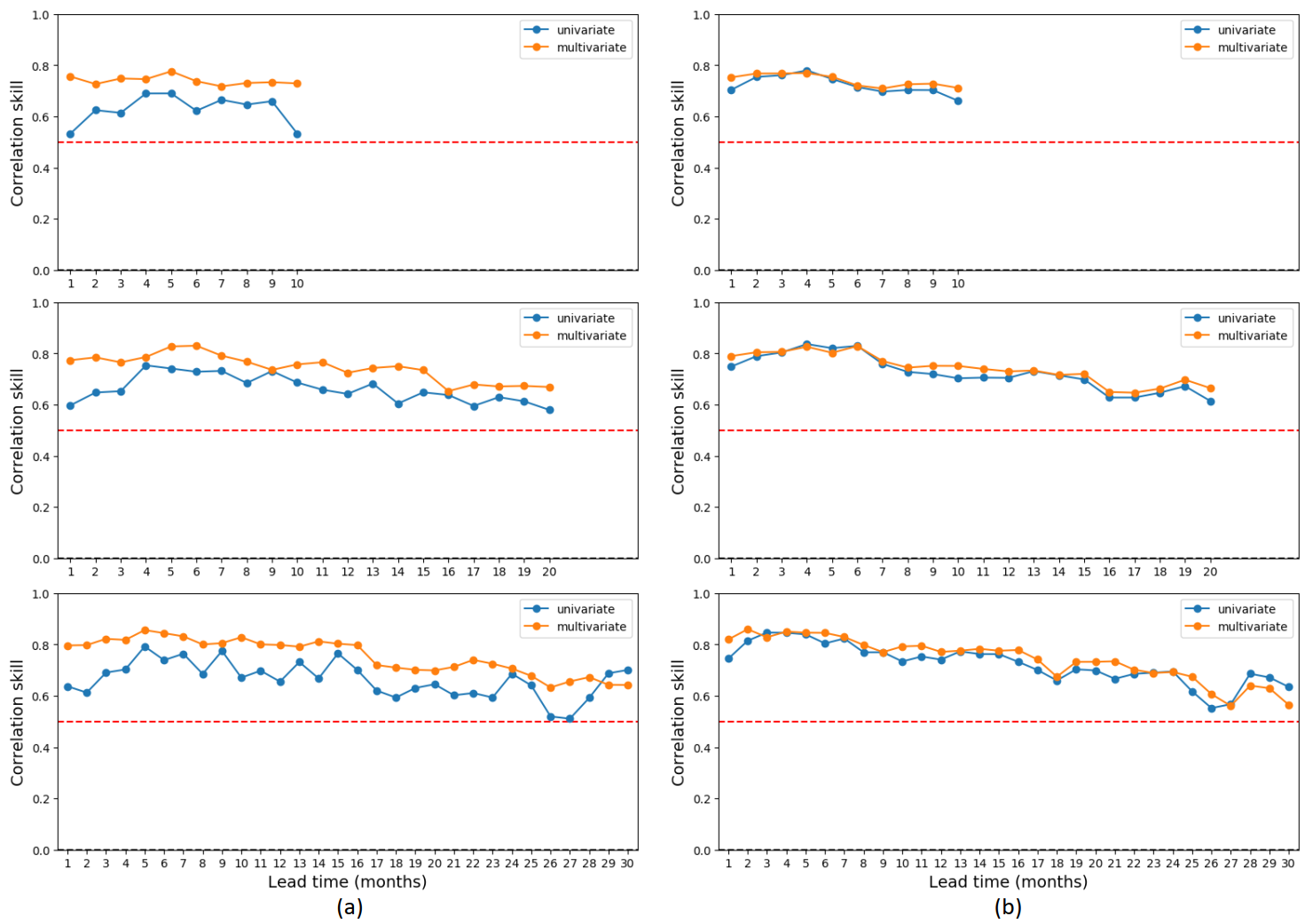} 
  \caption{Correlation skills of using the univariate (past ASI) and multivariate (past ASI plus Niño indices) predictors as inputs of the LSTM-based model, (a) the length of input sequences is 3;  (b) the length of input sequences is 6.
  }
  \label{fig2}
\end{figure}

\section{Experiments}

\paragraph{Evaluation} To evaluate the forecast skill of the LSTM-based model, we adopt the temporal anomaly correlation coeﬃcient $C$ in \citep{ham2019deep}.  $C$ is a function of the forecast lead months $l$ and measures the linear correlation between the actual and the predicted values. In addition, we also use Pearson correlation coefﬁcients (Corr) and Mean absolute percentage error (MAPE) to compare the true and the predicted time series during the validating period.

\begin{figure}
  \centering
  \includegraphics[width=\linewidth]{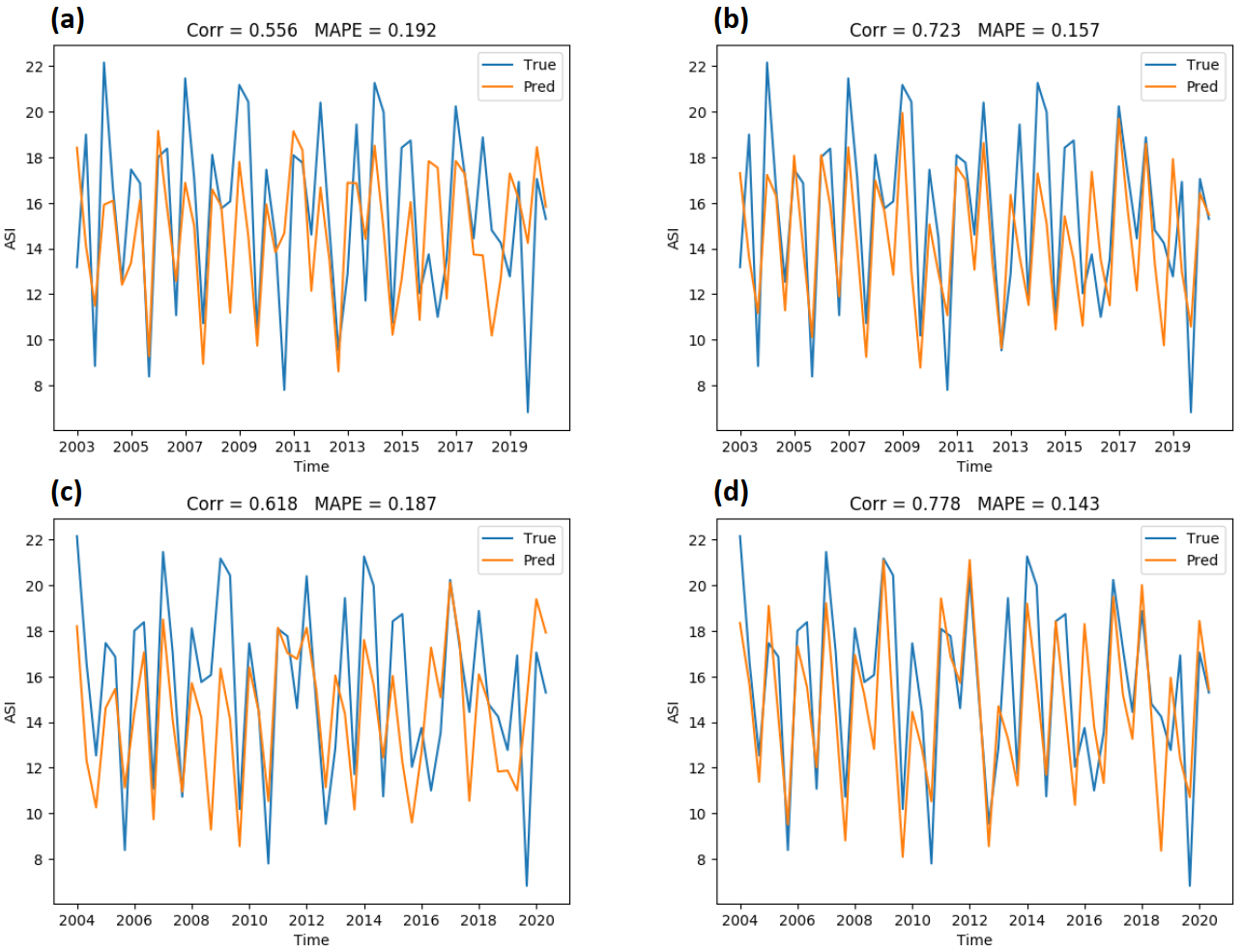} 
  \caption{Time series of the true and the predicted DJF ASI in southern China: using univariate input with a sequence length of 6 (a) or 9 (c); using multivariate inputs with a sequence length of 6 (b) or 9 (d). 
  }
  \label{fig3}
\end{figure}

\paragraph{Results}  
Figure~\ref{fig2} shows the correlation skill of the LSTM-based model at lead times of 10, 20, and 30 months (the first, second, and third row) using the univariate (blue line) and multivariate inputs (orange line) with the input sequence length of 3 (a) and 6 (b) respectively. Both curves of univariate and multivariate inputs in Figure~\ref{fig2} show a downward trend along with increasing lead time, which is reasonable.  Figure~\ref{fig2} (a) indicates that multivariate inputs achieve significantly better forecast skills than univariate input at lead times of 10, 20, and 30 months (both lines almost above 0.5), while Figure~\ref{fig2} (b) shows the univariate input can achieve an approximate forecast performance to multivariate inputs when the input sequence length is 6. Niño indices have been proven to be beneficial to improve the forecast ability of wintertime ASI when the input sequence length is short.

Furthermore, Figure~\ref{fig3} shows the true and predicted DJF ASI during the validation period using univariate and multivariate inputs with different input sequence lengths. We can observe that multivariate inputs achieve higher correlation coefficients and lower MAPE than univariate input under the same sequence length by comparing Figure~\ref{fig3} (a) and (b), (c) and (d). Finally, multivariate inputs with a sequence length of 9 achieve the best predictive performance with a high correlation coefficient of 0.778 and a low MAPE of 0.143.

\section{Conclusion and future work}
In this work, we have explored leveraging ocean memory effects to achieve long-lead forecasts of wintertime ASI. We first find negative correlations between autumn Niño indices and wintertime ASI in southern China, indicating the prospects for a prewinter forecast. Based on these correlations, an LSTM-based forecast model has been developed. Experimental results show that Niño indices are beneficial to help improve the forecast performance of wintertime ASI, especially when the input sequence is short. In future work, we will conduct more investigations: (1) try more powerful machine learning models, such as Transformer and ConvLSTMs, to further enhance the prediction accuracy; (2) directly use SST anomalies as predictors which may provide global and more useful information for long-lead forecasts.

\bibliography{iclr2023_conference}
\bibliographystyle{iclr2023_conference}   %{plain} %

% \appendix
% \section{Appendix}

\end{document}